\documentclass{article}
\title{Some simple physical facts about the collapse of the Twin Towers}
\author{Juan Betancort Rijo}
\begin{document}
\maketitle

To understand the collapse of these towers, it is useful to consider the following simple mechanical
problem: take a pile of $n$ equal blocks of height $s$, put on top of each other separated by empty gaps
of height $h$ containing only the supporting beams. Now, if the beams inmediately below the uppermost block
are suddenly blown off, it would start falling with acceleration $g$ and zero initial velocity. After 
falling a height $h$, it collides with the next block at a speed $\sqrt{2gh}$. If the corresponding
kinetic energy of the block is large compared with with the energy required for bending or breaking the
beams, conservation of linear momentum imply (completely inelastic collision) both blocks moving together
with speed $\sqrt{2gh}/2$ and acceleration $g$. After falling a height $h$, both blocks collide with the
next one at certain velocity, thereby moving the three blocks together at $2/3$ of this speed, and so on.
The lower surface of the falling pile we may call collapse front and its motion is a free fall in between
collisions and discontinuous (considering the collisions instantaneous) changes of velocity and position
(sadden jump by an amount $s$) at collisions. However, for $n$ large, it is interesting to consider the 
average motion of the collapsing front. In the limit $s \ll h$ (the correction for finite $s/h$ being 
trivial) it may easily be shown that this average motion is uniformly accelerated with acceleration $g/3$. 
More generally, taking into account the height of the blocks and the energy absorbed by the beams, one 
finds:
\begin{equation}
u(t) = \frac{1}{6} g \left( 1-\frac{s}{h+s} \right)^{-1}(1-f) t^2,
\end{equation}
where $u(t)$ is the position of the front from its initial position (the lower surface of the upper block),
$f$ is the ratio between the total energy absorbed by the beams and the potential energy of the pile and 
$t$ is the time elapsed from the start of the collapse.

Having mass in between the blocks (i.e. in the form of walls) does not change the nature of the collapse. 
If these masses move along with the blocks or are blown away at collision carrying no momentum, the 
acceleration is still $g/3$ (for $s \to 0$). If the fraction of momentum carried away by the falling wall 
is $F$, the collapse is an accelerated motion with acceleration $g(1-F)/3$. 

If the beams blown away were
those at the bottom, the pile would collapse (for $f \to 0$) with acceleration $g$. In this case, the 
collapsing front is static at the ground (for $s \to 0$, moving slightly upward for finite $s$).

Now consider the case when the beams blown away are those above the m-th block. In this case, assuming
$s=f=F=0$ (corrections being trivial) it may be shown that, before the uppermost block hit it, the front
falls with acceleration $3g/5$. The $n-m$ blocks above the front fall together with acceleration $g$ until
the uppermost block hit the front. At this time, the acceleration (average) changes suddenly to $7g/25$
and grows asimptotically to $g/3$.

However, for this mode of collapse to apply, the blocks must be completely incompressible, so that the
acceleration of the upper and lower surfaces of the collapsed pile experience not only the same mean
acceleration, but also the same very strong upwardly directed acceleration at collisions. In this case and 
assuming that the ratio between the actual statical charge supported and the maximum possible charge, $c$,
is constant all through the pile, the beams just over the upper surface of the collapsed pile are
equally likely to break as those below. On the other hand, when the blocks have a finite compressibility, 
the ratio between the statical charge and the dinamical charge at collisions is smaller above the upper
surface. In this case, after the collapsed pile have grown to some thickness, it provides a sufficient
cushion for the upper part to fall bodily resting on the collapsed pile. Then, the acceleration of the
collapse front is again $g/3$. The upper block falls with the same acceleration at a constant distance
from the front. The tremor of the collapse may cause the collapse from time to time of the block just
above the collapsing pile, but this results in constant average speed for the motion of the upper block
with respect to the front.

If the blocks break into irregular fragments, they would roll down from the top of the collapsed pile.
This mode of mass loss is different from the one considered previously and do not result in uniform 
acceleration. Asimptotically the collapsed pile approachs a maximum thickness, the acceleration decreases
and at some stage the beams will be strong enough to halt the collapse.

For the north tower, we have $s \approx 0.3$ m, $h \approx 3.3$ m, the initial position of the collapsing
front is about 370 m above the ground. Assuming $c=5$ constant all through the pile, it may be shown that
$f \leq 0.02$ \footnote{Using for steel: \\
- maximum pressure - 10$^8$ N/m$^2$\\ 
- maximum stress - 16$\times$10$^8$ N/m$^2$\\
- maximum strain - 0.004}. With $F=0$, the time for the collapse front to hit the
ground given by Eq. (1) is 14.56 s. Since the actual time seems to be around 17 s, this imply that the 
fraction of
mass blown away from the pile as the walls are squashed is somewhat above $1/4$ (assuming no matter fall 
from the back of the collapsed pile).
\end{document}